\documentclass{vldb}

\usepackage{graphicx}
\usepackage{hyperref}
\usepackage{balance}  
\usepackage[usenames,dvipsnames,svgnames,table]{xcolor}

\newcommand{\topkmd}{\textit{Top-k m\textsuperscript{th}} }
\newcommand{\topkfirstd}{\textit{Top-k 1\textsuperscript{st}} }
\begin{document}
	
	
	\title{Effective and Efficient\\ Variable-Length Data Series Analytics}

	
	
	%
	%
	%
	%
	
	\numberofauthors{1} 
	
	\author{
		%
		%
		\alignauthor
		Michele Linardi\\
		Supervised by: Themis Palpanas\\
		\affaddr{LIPADE, Universit{\'e} de Paris}\\
		\email{michele.linardi@parisdescartes.fr}
	}

	\maketitle
	
	\begin{abstract}
		In the last twenty years, data series similarity search has emerged as a fundamental operation at the core of several analysis tasks and applications related to data series collections. Many solutions to different mining problems work by means of similarity search. In this regard, all the proposed solutions require the prior knowledge of the series length on which similarity search is performed. In several cases, the choice of the length is critical and sensibly influences the quality of the expected outcome. Unfortunately, the obvious brute-force solution, which provides an outcome for all lengths within a given range is computationally untenable. 
		In this Ph.D. work, we present the first solutions that inherently support scalable and variable-length similarity search in data series, applied to sequence/subsequences matching, motif and discord discovery problems.
		The experimental results show that our approaches are up to orders of magnitude faster than the alternatives. They also demonstrate that we can remove the unrealistic constraint of performing analytics using a predefined length, leading to more intuitive and actionable results, which would have otherwise been missed.

	\end{abstract}

	\section{Introduction}
	
	Data series (i.e., ordered sequences of points) are one of the most common data types\footnote{If the dimension that imposes the ordering of the sequence is time then we talk about time series. Though, a series can also be defined over other measures (e.g., angle in radial profiles in astronomy, mass in mass spectroscopy in physics, position in genome sequences in biology, etc.). We use the terms \emph{data series}, \emph{time series}, and \emph{sequence} interchangeably.}, present in almost every scientific and social domain (such as meteorology, astronomy, chemistry, medicine, neuroscience, finance, agriculture, entomology, sociology, smart cities, marketing, operation health monitoring, human action recognition and others)~\cite{DBLP:journals/sigmod/Palpanas15}. 
	
	Once the data series have been collected, the domain experts face the arduous tasks of processing and analyzing them~\cite{KostasThemisTalkICDE, DBLP:conf/edbt/GogolouTPB19} in order to gain insights, e.g., by identifying similar patterns, and performing classification, or clustering.
	A core operation that is part of all these analysis tasks is similarity search, which has attracted lots of attention because of its importance~\cite{DBLP:journals/pvldb/EchihabiZPB18}.
	Nevertheless, all existing scalable and index-based similarity search techniques are restricted in that they only support queries of a fixed length, and they require that this length is chosen at index construction~\cite{DBLP:conf/vldb/PalpanasCGKZ04,shieh2008sax,DBLP:conf/icdm/CamerraPSK10,DBLP:journals/pvldb/WangWPWH13,ZoumpatianosIP15,DBLP:journals/vldb/ZoumpatianosIP16,DBLP:conf/icdm/YagoubiAMP17,DBLP:conf/bigdataconf/PengFP18,coconut}. 
	The same observation holds for techniques proposed to discover motifs~\cite{DBLP:conf/icde/LiUYG15} and discords (i.e., anomalous subsequences)~\cite{DBLP:journals/kais/YankovKR08}: they all assume a fixed sequence length, which has to be predefined.
	
	Evidently, this is a constraint that penalizes the flexibility needed by analysts, who often times need to analyze patterns of slightly different lengths (within a given data series collection)~\cite{DBLP:journals/kais/KadiyalaS08,914838,DBLP:conf/kdd/RakthanmanonCMBWZZK12,VALMOD,VALMOD_DEMO}. 
	For example, in the \emph{SENTINEL-2} mission data, oceanographers are interested in searching for similar coral bleaching patterns\footnote{\scriptsize\url{http://www.esa.int/Our_Activities/Observing_the_Earth}} of different lengths; at Airbus\footnote{\scriptsize\url{http://www.airbus.com/}} engineers need to perform similarity search queries for patterns of variable length when studying aircraft takeoffs and landings~\cite{Airbus}; and in neuroscience, analysts need to search in Electroencephalogram (EEG) recordings for Cyclic Alternating Patterns (CAP) of different lengths (duration), in order to get insights about brain activity during sleep~\cite{ROSA1999585}.
	
	In our work, we focus on three core problems that are based on similarity search: subsequence matching, and motif and discord discovery, organized under the ULISSE and MAD methods:
	
	1. ULISSE (ULtra compact Index for variable-length Similarity SEarch in data series) is the first indexing technique that supports variable-length subsequence matching for non Z-normalized and Z-normalized data series~\cite{ULISSEVldb,ULISSEJournal,ulisseICDE}.
	
	2. MAD (Motif and Discord discovery framework) implements two novel algorithms for variable-length motif and discord discovery in large data series~\cite{VALMOD,MAD,VALMOD_DEMO}.
	
	\section{Variable-Length Analytics}
	
	In this section, we describe our proposed approaches to the aforementioned problems. In the next part we describe the notions and the elements used in our solutions. 
	
	\noindent{\bf Preliminaries.} Let a data series $D = d_1$,...,$d_{|D|}$ be a sequence of numbers $d_i \in  \mathbb{R}$, where $i \in \mathbb{N}$ represents the position in $D$. 
	We denote the length, or size of the data series $D$ with $|D|$.
	The subsequence $D_{s,\ell}$=$d_s$,...,$d_{s+\ell-1}$ of length $\ell$, is a contiguous subset of $\ell$ points of $D$ starting at offset $s$, where $1 \leq s \leq |D|$ and $ 1 \leq \ell \leq |D|$.
	A subsequence is itself a data series.
	A data series collection, $C$, is a set of data series. 
	We say that a data series $D$ is Z-normalized, denoted $D^{n}$, when its mean $\mu$ is $0$ and its standard deviation $\sigma$ is $1$. 
	Z-normalization is an essential operation in several applications, because it allows similarity search irrespective of shifting and scaling~\cite{DBLP:conf/kdd/RakthanmanonCMBWZZK12}.
	The Piecewise Aggregate Approximation (PAA) of a data series $D$, $PAA(D) = \{p_{1},...,p_{w}\}$, represents $D$ in a $w$-dimensional space by means of $w$ real-valued segments of length $s$, where the value of each segment is the mean of the corresponding values of $D$~\cite{Keogh2000}. 
	We denote the first $k$ dimensions of $PAA(D)$, ($k \le w$), as  $PAA(D)_{1,..,k}$.
	
	The $iSAX$ representation of a data series $D$, denoted by $iSAX(D,w,|alphabet|)$, is the representation of $PAA(D)$ by $w$ discrete coefficients, drawn from an alphabet of cardinality $|alphabet|$~\cite{shieh2008sax}. 
	The main idea of the $iSAX$ representation, is that the real-value space may be segmented by $|alphabet|-1$ breakpoints in $|alphabet|$ regions, which are labeled by discrete symbols (e.g., with $|alphabet|=4$ the available labels may be $\{00,01,10,11\}$). 
	
	\subsection{Subsequence Matching}
	The subsequence matching problem is defined as follows: 
	
	Given a data series collection $C=\{D^{1},...,D^{C}\}$, a series length range $[\ell_{min},\ell_{max}]$, a query data series $Q$, where $\ell_{min} \le |Q| \le \ell_{max}$, and $k \in \mathbb{N}$, we want to find the set $R=\{D^{i}_{o,\ell}| D^{i} \in C \wedge \ell=|Q| \wedge (\ell+o-1) \le |D^{i}|\}$, where $|R|=k$. We require that $ \forall D^{i}_{o,\ell} \in R$ $\nexists D^{i'}_{o',\ell'}$ $s.t.$ $dist(D^{i'}_{o',\ell'},Q) < dist(D^{i}_{o,\ell},Q)$, where $\ell'=|Q|$, $(\ell'+o'-1) \le |D^{i'}|$ and $D^{i'} \in C$. We informally call $R$, the $k$ \textit{nearest neighbors} set of $Q$. Given two generic series of the same length, namely $D$ and $D'$ the function $dist(D,D')$ can be Euclidean Distance or Dynamic Time Warping.

	\noindent{\bf Variable Length Subsequences.} In a data series, when we consider contiguous and overlapping subsequences of different lengths within the range  $[\ell_{min},\ell_{max}]$,we expect the outcome as a bunch of similar series, whose differences are affected by the misalignment and the different number of points. 
	Given a data series $D$, and a subsequence length range $[\ell_{min},\ell_{max}]$, we define the master series as the subsequences of the form $D_{i,min(|D|-i+1,\ell_{max})}$, for each $i$ such that $1 \le i \le |D|-(\ell_{min}-1)$, where $1 \le \ell_{min} \le \ell_{max} \le |D|$.
	
	We observe that for any master series of the form $D_{i,\ell'}$, we have that $PAA(D_{i,\ell'})_{1,..,k} = PAA(D_{i,\ell''})_{1,..,k}$ holds for each $\ell''$ such that $\ell''\ge \ell_{min}$, $\ell''\le \ell' \le \ell_{max}$ and $\ell',\ell'' \% k = 0$.
	
	Therefore, by computing only the $PAA$ of the master series in $D$, we are able to represent the $PAA$ prefix of any subsequence of $D$. 
	When we zero-align the $PAA$ summaries of the master series, we compute the minimum and maximum $PAA$ values (over all the subsequences) for each segment: this forms what we call an \textit{Envelope}. 
	(When the length of a master series is not a multiple of the $PAA$ segment length, we compute the $PAA$ coefficients of the longest prefix that is multiple of a segment.)
	We call \textit{containment area} the space in between the segments that define the \textit{Envelope}.
	
	\noindent{\bf PAA Envelope.} We formalize the concept of the \textit{Envelope}, introducing a new series representation.
	We denote by $L$ and $U$ the $PAA$ coefficients, which delimit the lower and upper parts, respectively, of a containment area. 
	Furthermore, we introduce a parameter $\gamma$, which permits to select the number of master series we represent by the \textit{Envelope}. 
	We refer to it using the following signature: $paaENV_{[D,\ell_{min},\ell_{max},a,\gamma,s]} = [L,U]$.
	It delimits the containment area generated by the $PAA$ coefficients of the master series.
	
	\noindent{\bf Indexing the Envelopes.} Given a $paaENV$, we can translate its $PAA$ extremes into the corresponding iSAX representation: $uENV_{ paaENV_{[D,\ell_{min},\ell_{max},a,\gamma,s]}}=[iSAX(L),iSAX(U)]$, where $iSAX(L)$ ($iSAX(U)$) is the vector of the minimum (maximum) $PAA$ coefficients of all the segments corresponding to the subsequences of $D$.
	The \textit{Envelope} $uENV$ represents the principal building block of the \textit{ULISSE} Index.
	In details, \textit{ULISSE} is a tree structure, where each internal node stores the \textit{Envelope} $uENV$ representing all the sequences in the subtree rooted at that node. 
	Leaf nodes contain several \textit{Envelopes}, which by construction have the same $iSAX(L)$. On the contrary, their $iSAX(U)$ varies, since it get updated with every new insertion in the node. Each \textit{Envelope} in leaf nodes point the the represented sequences in the original data series collection.

	\noindent{\bf Approximate Subsequence Matching.} Subsequence matching performed on \textit{ULISSE} index relies on the $mindist_{ULiSSE}()$ lower bounding function to prune the search space. 
	This allows to navigate the tree in order, visiting first the most promising nodes. As soon as a leaf node is discovered, we can load the raw data series pointed by the \textit{Envelopes} in the leaf.
	Each time we compute the true Euclidean or DTW distance between the series in a leaf, the best-so-far distance (bsf) is updated, along with a vector containing the $k$ best matches, where $k$ refers to the $k$ nearest neighbors.
	Since priority is given to the most promising nodes, we can terminate our visit, when at the end of a leaf visit the $k$ bsf's have not improved.
	
	\noindent{\bf Exact Subsequence Matching.}
	Note that the approximate search described above may not visit leaves that contain answers better than the approximate answers already identified, and therefore, it will fail to produce exact, correct results.
	
	The exact nearest neighbor search algorithm we propose finds the $k$ sequences with the absolute smallest distances to the query. In this case, the search algorithm may visit several leaves: the process stops after it has either visited, or pruned (when the lower bounding distance to the node is greater than the bsf) all the nodes of the index, guaranteeing the correctness of the results.
	
	
	\noindent{\bf Experiments.} To evaluate \textit{ULISSE}, we used synthetic and real data (but in the interest of space we only report results with the synthetic data). We record the average \textit{CPU time}, \textit{disk I/O} (time to fetch data from disk (Total time - CPU time)), for \textit{100} queries, extracted from the datasets with the addition of Gaussian noise. 
	We compare \textit{ULISSE} with \textit{UCR suite}~\cite{DBLP:conf/kdd/RakthanmanonCMBWZZK12} the non index-based state-of-the-art technique for answering similarity search queries. 
	Concerning the competitor indexing techniques, the state-of-the-art is the Compact Multi Resolution Index~\cite{DBLP:journals/kais/KadiyalaS08} \textit{CMRI}.
	
	\begin{figure}[tb]
		\includegraphics[trim={0cm 14cm 10cm 3cm},scale=0.55]{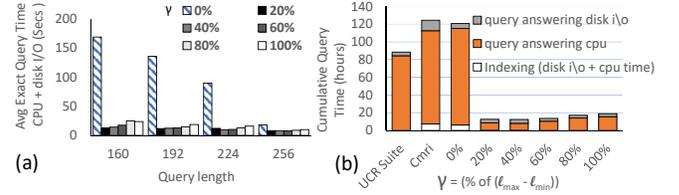}
		\caption{Query answering time performance, varying $\gamma$ on non Z-normalized data series. \textit{(a)} \textit{ULISSE} average query time (CPU + disk I/O). \textit{(b)} \textit{ULISSE} average query disk I/O time. \textit{(b)} Comparison of \textit{ULISSE} to other techniques (cumulative indexing + query answering time).}
		\label{1Exp}
		\vspace*{-0.5cm}
	\end{figure}
	
	In Figure~\ref{1Exp}, we present results for subsequence matching queries on $\textit{ULISSE}$ when we vary~$\gamma$, ranging from \textit{0} to its maximum value in this dataset, i.e., $\ell_{max}-\ell_{min}$. 
	In Figure~\ref{1Exp}, we report the results concerning non Z-normalized series.
	We observe that grouping contiguous and overlapping subsequences under the same summarization (\textit{Envelope}) by increasing $\gamma$, affects positively the performance of index construction, as well as query answering. 
	
	The latter may seem counterintuitive, since inserting more master series into a single \textit{Envelope} is likely to generate large containment areas, which are not tight representations of the data series.
	On the other hand, it leads to an overall number of \textit{Envelope} that is several orders of magnitude smaller than the one for $\gamma=0\%$, where only a single master series is represented by each \textit{Envelope}.

	\subsection{Motif and Discord Discovery}
	Motif and Discord are data mining primitives that represent frequent and rare (anomalous) patterns, respectively. Given a data series D, they are defined as follows:
	\begin{itemize}
		\item Data series motif: $D_{a,\ell}$ and $D_{b,\ell}$ is a motif pair iff $dist(D_{a,\ell},D_{b,\ell})  \le dist(D_{i,\ell},D_{j,\ell})$
		$\forall i,j \in[1,2,...,|D|-\ell+1]$, where $a \neq b$ and $i \neq j$, and dist is a function that computes the z-normalized Euclidean distance between the input subsequences.
		\item Data series discord: We call the $k$ subsequences of $D$, with the $k$ largest distances to their $m^{th}$ Nearest Neighbor (according the Euclidean distance), the \topkmd discords.
	\end{itemize}
	
	\noindent{\bf Variable length motif and discord discovery.} We provide solutions to the following problems:
	\begin{itemize}
		\item Variable-Length Motif Discovery:  Given a data series $D$ and a subsequence length-range $[\ell_{min},...,\ell_{max}]$, we want to find the data series motif pairs of all lengths in $[\ell_{min},...,\ell_{max}]$, occurring in $D$. 
		\item Variable-Length \topkmd Discord Discovery: Given a data series $D$, a subsequence length-range $[\ell_{min},...,\ell_{max}]$ and the parameters $a,b \in \mathbb{N^{+}}$ we want to enumerate the \topkmd discords for each $k \in \{1,..,a\}$ and each $m \in \{1,..,b\}$, and for all lengths in $[\ell_{min},...,\ell_{max}]$, occurring in $D$.
	\end{itemize}
	
	\noindent{\bf Fixed length motif and discord discovery.} 
	The state-of-the art algorithm for fixed length motif and discord discovery~\cite{DBLP:conf/icdm/YehZUBDDSMK16} requires the user to define the length of the desired motif or discord. This mining operation is supported by computation of the \textit{Matrix profile}, which is a meta data series storing the z-normalized Euclidean distance between each subsequence and its nearest neighbor. The Matrix profile does not only derive the motif, but also ranks and filters out the other pairs, giving also a convenient and graphical representation of their occurrences and proximity.
	Unfortunately, this technique comes with an important shortcoming: it does not provide an effective solution for trying several different motif lengths.
	Therefore, the analyst is forced to run the algorithm using all possible lengths in a range of interest, and rank the various motifs discovered, picking eventually the patterns that contain the desired insight. 
	Clearly, this possibility is not optimal for at least two reasons: the scalability, since finding motif of one fixed length takes $O(|D|^2)$ time, and also because it does not provide an effective way to compare motifs of different lengths.\\
	\noindent{\bf MAD Framework.} Our framework for Variable Length Motif and Discord Discovery (MAD) works by applying an \textit{incremental computing} strategy, which aims to prune unnecessary distance computations for larger motif and discord lengths.
	Hence, given a data series $D$, we compute the Matrix profile using the smallest subsequence length, namely $\ell_{min}$, within a specified input range $[\ell_{min},\ell_{max}]$.
	The key idea of our approach is to minimize the work that needs to be done for succeeding subsequence lengths ($\ell_{min}+1$, $\ell_{min}+2$, $\ldots$, $\ell_{max}$).
	

	
	\noindent{\bf Matrix Profile Computation.} We start the computation of the Matrix profile, considering all the contiguous subsequences of length $\ell_{min}$, computing for each one the \textit{Distance profile} in $O(|D|)$ time. This latter is a vector that contains the z-normalized Euclidean distances between a fixed subsequence and all the other in $D$ (excluding trivial matches). 
	
	\noindent{\bf Lower Bound Subsequences of Different Length.} We moreover introduce a new lower bounding distance~\cite{VALMOD}, which lower bounds (is always smaller than) the true Euclidean distances between subsequences longer than $\ell_{min}$. 
	We initially compute this lower bound using the true Euclidean distances computation of subsequences with length $\ell_{min}$. For the larger lengths, we update the lower bound, considering only the variation generated by the trailing points in the longer subsequences. This measure enjoys an important property: if we rank the subsequences according to this measure, the same rank will be preserved along all the lower bound updates for the subsequences of greater length. 
	We exploit this property, in order to prune computations. 
	
	\noindent{\bf Pruning the Search Space.} 
	Once we compute motif and discords, with length greater than $\ell_{min}$, instead of computing from scratch each distance profile, we update the true distances (in constant time) of the subsequences that have the $p$ smallest lower bounding distances (computed in the previous step). These distances form what we call \textit{partial distance profile}.
	In each partial distance profile, we also update the lower bound. 
	After this operation, we may have two cases: if in a new computed distance profile the minimum true distance (\textit{minDist}) is shorter than the maximum lower bound (\textit{maxLB}), we know that no distances among those not computed can be smaller than minDist.
	In this case, a partial distance profile becomes a \textit{valid distance profile}.
	On the other hand, when \textit{maxLB} is smaller than \textit{minDist}, this latter is not guaranteed to be the nearest neighbor distance. For discord discovery, we need to test this condition for the $m$ smallest true distances in the partial distance profile. In this case a valid (partial) distance profile must contain the true $m^{th}$ best match distances, which are smaller than, or equal to  \textit{maxLB}. 
	
	\noindent{\bf Exact Motif and Discord Discovery.} Once the partial distance profiles are computed, we pick the absolute smallest lower bounding value from all the non-valid distance profiles, namely \textit{minLBAbs} (if any).
	Therefore, the global minimum (true) distance of all the valid (partial) distance profiles, which is smaller than \textit{minLBAbs} is guaranteed to be the distance between the motif pair subsequences.
	Symmetrically, we consider the valid (partial) distance profiles to find the true $m^{th}$ best match distances, which are the greatest nearest neighbor distances that are larger than \textit{maxLBAbs}. This latter is the largest lower bounding distance of the non-valid distance profiles.
	
	In the motif discovery task, if no nearest neighbor distance is smaller than \textit{minLBAbs}, we recompute only the distance profiles that have the \textit{maxLB} distance smaller than the smallest true distance computed.
	
	On the other hand, for discord discovery, if no true nearest neighbor distances are found we need to iterate the non valid (partial) distance profiles, which contain the \textit{maxLB} distance greater than the largest $m^{th}$ best match distance.     
	
	We keep extracting in this manner the motif and the discord subsequences of each length, until $\ell_{max}$.

\begin{figure}[tb]
		\centering
		\includegraphics[trim={2cm 15.5cm 2cm 0cm},scale=0.4]{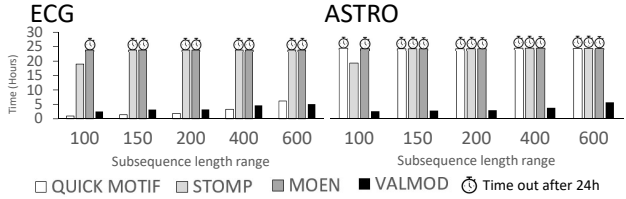}
		\caption{ Time over motif length ranges (default $\ell_{min}$=\textit{1024}, data series length= \textit{0.5M}.}  
		\label{Scalability1}
		\vspace*{-0.5cm}
	\end{figure}

	\noindent{\bf Motif Discovery Experimental Evaluation.} To benchmark the MAD framework, we used several different real datasets.
	Concerning the motif discovery problem, the competitors we considered are: QUICKMOTIF~\cite{DBLP:conf/icde/LiUYG15}, STOMP~\cite{DBLP:conf/icdm/YehZUBDDSMK16}, and MOEN~\cite{DBLP:journals/kais/MueenC15}.
	We report in Figure~\ref{Scalability1} a sample of the experiments we conducted (detailed experimental results on several datasets are reported elsewhere~\cite{VALMOD}). 
	Here, we show the results of MAD, which finds motifs in different real datasets. In the plots, we report the total execution time varying motif length ranges.
	From this experiment, we observe that VALMOD maintains a good and stable performance across datasets and parameter settings, quickly producing results, even in cases where the competitors do not terminate within a reasonable amount of time.
	
\begin{figure}[tb]
		\includegraphics[trim={1cm 17cm 3cm -1cm},scale=0.45]{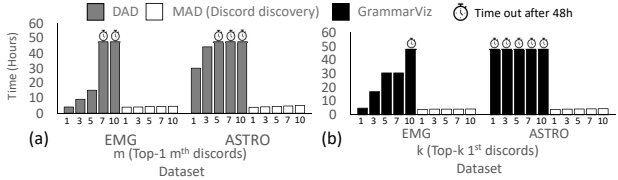}
		\caption{\text{(a)} $Top-1$ $m^{th}$ discords discovery, and \text{(b)} $Top-k$ $1^{st}$ discords discovery time performance.}  
		\label{discordsPlots}
		\vspace*{-0.5cm}
	\end{figure}

	\noindent{\bf Discord Discovery Experimental Evaluation.} We identify two state-of-the-art competitors to compare to our approach, the Motif And Discord (MAD) framework.
	The first one, DAD (Disk aware discord discovery)~\cite{DBLP:journals/kais/YankovKR08}, implements an algorithm suitable to enumerate the \emph{fixed-length} $Top-1$ $m^{th}$ discords.
	The second approach, GrammarViz~\cite{DBLP:conf/edbt/Senin0WOGBCF15}, is the most recent technique, which discovers \topkfirstd discords.
	In Figure~\ref{discordsPlots}.(a) we report the results of $Top-1$ $m^{th}$ discord discovery, varying $m$. We note that MAD gracefully scales over the number of discords to enumerate and is up to one order of magnitude faster than DAD. In Figure~\ref{discordsPlots}.(b), we show the result of $Top-k$ $1^{st}$ discords discovery. 
	Once again, MAD scales better over the number of discovered discords, as its execution time remains almost constant.
	A different trend is observed for GrammarViz, whose performance significantly deteriorates as $k$ increases.

\section{Conclusions}
Even though much effort has been dedicated for developping techniques for data series analytics, existing solutions for subsequence matching, motif and discord discovery are limited to fixed length queries/results.
In this Ph.D. work, we propose the first  scalable solutions to the variable-length version of these problems: \textit{ULISSE} is the first index that supports variable-length subsequence matching over both Z-normalized and non Z-normalized sequences~\cite{ULISSEVldb,ULISSEJournal,ulisseICDE}, while MAD is the first framework that implements variable-length motif and discord discovery~\cite{VALMOD,MAD,VALMOD_DEMO}.
	
	\balance
	

	\def\thebibliography#1{
		\section*{References}
		\scriptsize
		\list
		{[\arabic{enumi}]}
		{\settowidth\labelwidth{[#1]}
			\leftmargin\labelwidth
			\parsep 0pt                
			\itemsep 0pt               
			\advance\leftmargin\labelsep
			\usecounter{enumi}
		}
		\def\newblock{\hskip .11em plus .33em minus .07em}
		\sloppy\clubpenalty10000\widowpenalty10000
		\sfcode`\.=1000\relax
	}

	\bibliographystyle{abbrv}
	\bibliography{vldb_workshop}  
	
	
	
	%
	%
	%
	%
	
\end{document}